\documentclass[12pt]{article}
 \usepackage{epsfig} \usepackage{amsmath} \usepackage{amssymb}
 \usepackage{pifont} \usepackage{psfrag} \usepackage{graphicx}
\newcommand{\beq}{\begin{equation}}
\newcommand{\eeq}{\end{equation}}
\newcommand{\ns}{\normalsize} \newcommand{\scs}{\scriptsize}
  \newcommand{\fns}{\footnotesize}
  \newcommand{\veps}{\varepsilon}

\begin{document}
\begin{center}
 {\bf\large 60 years of Broken Symmetries in Quantum Physics} \\
 \bigskip
  {\ns\sf (From the Bogoliubov Theory of Superfluidity to the
  Standard Model) }
  \medskip

 {\small Shirkov D.V.}
  \end{center} \vspace{-2mm}
  \begin{flushright}
 {\it ``Phase transition in quantum system, as a rule, is \\
       accopanied by Spontaneous Symmetry Breaking'' } \\
  {\small Folklore \ of the middle of the  \ XX \ century} \end{flushright}

 \begin{abstract} A retrospective historical overview of the
 phenomenon of spontaneous symmetry breaking (SSB) in quantum
 theory, the issue that has been implemented in particle physics
 in the form of the Higgs mechanism. The main items are:\\
  -- The Bogoliubov's microscopical theory of superfluidity
  (1946);\\
  -- The BCS-Bogoliubov theory of superconductivity (1957);\\
  -- Superconductivity as a superfluidity of Cooper pairs
  (Bogoliubov - 1958);\\
  -- Transfer of the SSB into the QFT models (early 60s);\\
  -- The Higgs model triumph in the electro-weak theory
  (early 80s);\\
   The role of the Higgs mechanism and its status in the
  current Standard Model is also touched upon.
 \end{abstract}

 \section{\sf Introduction}
  Spontaneous symmetry breaking is a well-established term
 in quantum theory; its essence is simple. One has in mind a
 physical system that can be described by expressions
 (Lagrangian, Hamiltonian, equations of motion) obeying some
 symmetry, while a real physical state of the system
 corresponding to some partial solution of the equations of
 motion does not obey this symmetry. One meets such a case
 when the lowest of possible symmetrical states does not
 provide the system with absolute energy minimum and turns
 out to be unstable. A particular lowest state is not unique;
 a full collection of them forms a symmetric set. The real
 cause of symmetry breaking and transition of the
 system to some of the lowest non-symmetrical states usually
 turns out to be an arbitrary small asymmetrical perturbation.

 \begin{figure}[ht!] \begin{center}
 \includegraphics[scale=.90]{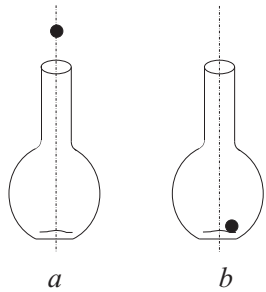}
 \caption{\small \ Simple mechanical system illustrating \
  spontaneous symmetry breaking: \ (a)\ initial state;\
 (b)\ final state.} \label{fig1ufn} \end{center}
 \end{figure}
  As a simple illustration take a system of an empty vessel with
 a convex bottom and a tiny massive ball. Let the vessel, which
 is a figure of revolution, stand vertically and the ball be
 located above it, just on the axis (Fig. 1а). The system is
 symmetric with  respect to rotation around the vertical axis.
 Let the ball fall down due to the force of gravity. Upon
 reaching the bottom, the ball will not stand at the center of
 convex surface and will roll down to some point at the periphery
 of the bottom (Fig.1б). Thus, the initial conditions are
 symmetrical, while the final state is not. \vspace{2mm}

 A more pithy example is the magnetized ferromagnet. The
 compass was known to the ancient Chineses, but only in the
 beginning of XVIII century an Oxford professor of astronomy
 John Keill\cite{keill} noticed that heating destroys magnetic
 property \footnote{The quotation is given with the
 orthography of the original.}: \\
 {\it ``...
 if a Loadstone be put into the Fire, insomuch that the internal
 Structure of its Parts be changed or wholly destroyed, then it
 will lose all its former Virtue, and will scarce differ from
 other Stones.''} 

  A systematical study of thermal properties of magnetic
 substances was undertaken by Pierre Curie, who discovered a
 sharp decrease of magnetization as the temperature
 approached the critical value, now called the Curie point.
 Above the critical temperature ferromagnetism disappears.
 With decreasing temperature from the critical point the direction
 of magnetization may be changed to the opposite one if a
 ferromagnetic is placed in the external field opposite to the
 reference direction of magnetization and then this field is
 removed. Thus, ferromagnetic magnetization is related with two
 important notions. First, it is spontaneous symmetry breaking,
 as the external field may be chosen as weak as one wishes.
 Second, the value of magnetization is just the quantity that
 was called the order parameter in the Landau theory of phase
 transitions \cite{dau37} (see also pp 234-252 in \cite{dau-1}).
 This parameter is nonzero in the ferromagnetic region and
 continuously decreases to the critical point where it vanishes.

  The main subject will be exposed on the material of quantum
 statistics (superfluidity and superconductivity) with a smooth
 transition to quantum field theory, as far as the recent
 upgrading of interest in Spontaneous Symmetry Breaking (SSB)
 stems from  the quantum-field context. The previous contribution
 by Dremin, which plunged the audience in the bulk of technical
 details of future experiments at the Large Hadron Collider
 (LHC), reminded us of the \ ``Higgs expectations''. The latter
 are tightly related with SSB. \par
  Incidentally, in our exposition, we will mention two diverse
 and partially opposing one another ways of conceiving main
 ideas on the structure of the physical world. That is the
 ways of constructing the physical theory.

  The initial feeding material of our science, the data from
 observations, are to be systematized and understood. To put in
 order, one usually constructs a phenomenological model that is
 based on some physical idea, the model invested in a mathematical
 form, the form of a physical law. An important criterion of
 successfulness of the scheme and its grounds is not only  a
 reasonable correlation of the initial data, but possibility
 prediction of new effects with a clear-cut way of their
 implementation. This is a usual road of a phenomenologist, the
 way ``from a phenomenon to a theoretical scheme'' and backwards.

  Along with this, many important steps in the building of the
 physical theory are performed by another, a more speculative way.
 Remind Heraklit, unification of the celestial and terrestrial
 gravity, electricity and magnetism, as well as the recently
 discovered principle of \ {\it dynamics from symmetry} \ that made
 the foundation of the electro-weak theory and quantum
 chromodynamics.

 Adherents to this way of thinking, the people that try to start
 from  deep and profound ideas, from primary principles {\it
 ab initio}, are known as ``reductionists''\footnote{One implies
 the leading tendency to {\it reduce} the description,
 understanding of the bulk of an observed variety of events to a
 smll amount of simple notions and general principles.}. In
 statistical physics the latter, as a rule, are adherents to a
 microscopical approach\footnote{We quote the definition
 formulated by Bogoliubov in the 1958 paper \ {\it ``Basic
 principles of the theory of superfluidity and
 superconductivity''} \ \cite{Bog58d} (see also pp 297-309
 in \cite{nnVIII}) :
  \vspace{-1mm} \begin{quote}
 {\scs\it  The goal of macroscopical theory can be said as
 obtaining equations, similar to classical equations of
 mathematical physics that describe a majority of data
 related to macroscopical objects under study. \dots}
  \end{quote}  \vspace{-2mm} and then  \vspace{-2mm}
 \begin{quote}
 {\scs\it  In microscopical theory, a more profound problem is
 posed: to understand an intrinsic mechanism of the phenomena,
 in terms of quantum mechanics notions and equations. \dots \ \
 Here, in particular, one should also obtain relations between
 dynamical variables; relations that yield equations of
 macroscopical  theory. } \end{quote} }.

  At the same time, the reductionists comprise an overwhelming
 majority of founders of basic fundamentals of modern physics
 like the theory of relativity, quantum mechanics and the
 theory of quantum fields. \par

  Meanwhile, in our opinion, one should not exaggerate an
 opposition of these two modes of reflection. An important detail
 is that between equations, e.g., equations of classical mechanics
 or Maxwell equations in medium (plasma) and laws that describe
 a sequel of observed events for instance, laws of a planet
 motion or the Meissner law in a superconductor, there is a
 space, a logical gap. Just here the phenomenology works. Due
 to this, efforts of reductionists and phenomenologists, at the
 very end, supplement each other. Turn to examples. \smallskip

  In the early 30s, by heuristic reasonings, Fermi devised a
 four-fermionic Lagrangian for a weak nuclear force, initially
 with one coupling constant $G_F\,.$ The Fermi Lagrangian, with
 subsequent modifications, played an important role for
 understanding and regulating numerous data on lepton dynamics.
 The Fermi model modification of the mid50s included up to 10
 coupling parameters.\\
 More profound understanding of the weak interaction was achieved
 a quarter of a century later, in the
 Glashow--Salam--Weinberg (GSW) gauge theory of electro-weak
 interaction  with its massive vector $W\,$ and $Z\,$ bosons that
 appeared to be a ``missing link'' transmitters of forces between
 lepton currents. The origin of heavy masses ($\sim 90\,$GeV)
 of these particles is connected with SSB. The GSW theory is
 elegant and rather simple, being based on the new general
 principle \ ``dynamics from symmetry''. A transition to a deeper
 level reduced greatly the number of parameters.\smallskip

  In the year of 1941, quite soon after the experimental discovery
 of superfluidity, Lev Davidovich Landau \ ``just on the move''
 as it Kapitsa said, devised a pheno\-menological model
 \cite{dau41}  (see also pp 352-385 in \cite{dau-1} and
 \cite{dau67}) that described quite well some essential
 properties of HeII -- thermodynamics, kinetics and so on.

  The pith of the Landau's reasoning was the assumption of the
 dominating role of the collective quantum effect. Analysis at
 the microscopic level appeared five years later as a model of a
 weakly imperfect Bose gas, when Nikolaj Nikolaevich Bogoliubov
 proposed to treat atoms of HeII as weakly repulsing particles
 interacting with condensate. Here the key element consisted in
 the admission of that the condensate contained a macroscopically
 large number of helium atoms. That was the hypothesis that led
 to the elucidation of the nature of the Landau collective
 effect. In his paper \cite{nn47} (see also pp 108-112 in
 \cite{nnVIII} and \cite{nn67}), the famous $(u,v)\,$
 transformation was introduced that is tightly related with
 spontaneous breaking of phase symmetry responsible for the
 conservation of the number of particles.

  The third example, finally. The remarkable 1950 paper by
 Ginzburg and Landau\cite{G-L-50} (see also pp 126-152 in
 \cite{dau-1}) -- phenomenological description of the
 superconductivity by a specially devised, rather abstract,
 wave-like function $\Psi(\mathbf r)\;$ (the two-component order
 parameter) of the collective of the superconducting electrons.
  However, the understanding of the function $\Psi(\mathbf r)\;$
 physical content appeared 8-9 years later, after elaborating
 the Bardin-Cooper-Schrieffer and, particularly, Bogoliubov
 microscopical constructions, explicitly taking into account
 the interaction of electrons with the ion lattice vibrations.

  \section{\sf SSB in quantum statistics}

 \subsection{Superfluidity}
 The theory of superfluidity is a good example of interconnection
 between phenomenolo\-gical ideas and mathematical constructions.
 The original explanation of the phenomenon of superfluidity
 offered by Landau\cite{dau41} was based on the idea that at low
 temperatures the properties of liquid ${\rm^4He}\,$ were defined
 by collective excitations (phonons) rather than a quadratic
 spectrum of individual particle excitations. It follows from
 this assumption that in moving with velocity not exceeding a
 certain critical value it is impossible to slow down the liquid
 by transferring energy and momentum from the wall to individual
 atoms because a linear form of the phonon spectrum does not
 allow one to obey simultaneously the laws of energy and
 momentum conservation. The need for agreement between the form
 of the spectrum and the thermodynamic properties of liquid
 helium motivated Landau to introduce particular excitations,
 in addition to phonons, with a quadratic spectrum
 beginning with a certain energy gap, excitation, which he
 called rotons\footnote{See below Fig. 2(a) in which formulae
 (2.2) and (2.3) from paper \cite{dau41} are used.}.\\
 Bogoliubov's theory is based on a physical assumption that in
 weakly nonideal Bose gas there is a condensate akin to ideal
 Bose gas. The existence of the Bose condensate leads to a
 unique wave function of the whole system, i.e., collective
 effect. Therefore, the presence of even a weak interaction
 transforms single-particle excitations into the spectrum of
 collective excitations. To calculate this spectrum, Bogoliubov
 inferred that at low temperatures the Bose condensate contains
 a macroscopically large\footnote{Bogoliubov's intuitive guess
 got later a direct data support -- see papers
 \cite{condens1,condens1a,condens2}.}, of an order of Avogadro
 number $N_A\,,$ number of particles $N\,$ as a result of which
 matrix elements of the creation and annihilation operators of
 particles in the condensate are proportional to \ ``large'' \
 number $\sim \sqrt{N_0}\,,$ and the main contribution to the
 system dynamics comes from the processes of particle transition
 from the condensate to the continuous spectrum and back to the
 condensate.\par
 Following paper \cite{nn47}, we start with the second-quantized
 description of the system of Bose particles in the coordinate
 representation. The Hamiltonian of system with a pair
 interaction looks like
  \begin{equation}\label{H-coord}
 H= -\tfrac{\hbar^2}{2m}\int d\,x\,\Psi^*(x)\Delta \Psi(x)+
 \int d\,x\int d\,y\,\Psi^*(x)\Psi(x)\,V(x-y)\,\Psi^*(y)\Psi(y).
 \end{equation}
 Extraction of the condensate corresponds to a transition
 of the $\Psi\,$ function to the sum
 \begin{equation}\label{sdvig}
 \Psi(x)= C +\phi(x)\,\quad \Psi^*(x)= C +\phi^*(x)\eeq
 of the ``large constant''\  $C\,$ (containing an identity
 operator) and the \ ``small operator'' $\phi(x)\,.$ Since the
 Fourier transform of the constant is the Dirac delta function,
 in the discrete momentum representation
 \begin{equation}\label{fourier}
 \Psi(x)=\tfrac{1}{\sqrt{V}}\,\sum_k a_k\,
 e^{\tfrac{i\,(k\,x)}{\hbar}}\,\quad \phi(x)=
 \tfrac{1}{\sqrt{V}} \,\sum_{p\neq 0}\,
  b_p e^{\tfrac{i\,(p\,x)}{\hbar}},\,\quad \dots\end{equation}
 one can write
\begin{equation}\label{}
 a_k= a_0\,\delta_{k,0}\,c + \left[1-\delta_{k,0}\right]\,
 \delta_{k,p}\,b_p\,;\quad c= C/\sqrt{V} =\sqrt{N_0/V}\,,
\end{equation}
 where  $a_k\,,a^*_k\,\,$ and   $\,b_p\,,b_p^* \,$ are operators
  with Bose commutation relations
 $$ a_k\,a^*_q-a^*_q\,a_k =\delta_{k,q}\,;\quad
  b_p\,b^*_l-b^*_l\,b_p =\delta_{p,l}\,.$$

 Under the assumption of the decisive role of the condensate one
 can neglect terms responsible for an interaction of
 above-condensate atoms with each other.\par
 Then the total Hamiltonian of Bose gas in the momentum
 representation
  \begin{equation}\label{BogHam0}
 H_{B_0}=\sum_k\,T(k)\,a^+_k\,a_k+\sum_{k,q} v(k_1-k_2)\,
  a^+_{k_1}\,a_{k_2}\,a^+_{q_1}\, a_{q_2}\,\delta_{k_1-k_2
  \,,q_1-q_2}\,; \quad T(k)=\tfrac{k^2}{2m}\,,\eeq
 with the Fourier transform $\,v(k)>0 \,$ of the potential energy
 of weak pair repulsion of helium atoms\footnote{Summation is over
 3-dimensional discrete momentum space corresponding to the system
 final volume $V$ in the coordinate space. The three-dimensional
 Kronecker symbol is related with the three-dimensional delta
 function $\mathbb{\delta}\,$ by
 $ V\,\delta_{k,q}\to (2\,\pi)^3\,\mathbb{\delta}(k-q)\,$
 as $V\to\infty\,.$} results in the Bogoliubov Hamiltonian
 \cite{nn47} of the weakly nonideal Bose gas model\footnote{Here
 and below in subsection 2.1 \ ``Superfluidity'' \ momentum
 $p\,,$ in contrast with  $k,q\,,$ does not take zero value
 being referred only to above-condensate particles.}
  $$ H_{B_0}\to H_0+H_{B_1}\,;\quad H_0=
 v(0)\,{\mathbf N}_0^2/2V\,,\quad{\mathbf N}_0=a^+_0\,a_0\,,$$
 where  $\,{\mathbf N}_0 =a^+_0\,a_0$ is the particle number
 operator (i.e., occupation number) in the condensate, and
 \begin{equation}\label{BogHam1}
 H_{B_1}=\sum_{p \neq 0}\left\{ T(p)+\frac{{\mathbf N}_0\,
 v(p)}{V} \right\}\,b^+_p b_p+
\frac{1}{2V}\sum_{p\neq 0}v(p)\{b^+_pb^+_{-p}a_0 a_0+
 a^+_0\,a^+_0\,b_p b_{-p}\}\,.\eeq
 The second sum describes particle transitions from the
 condensate and back, i.e., produc\-tion of pairs with zero total
 momentum from the condensate and their annihilation.\par

 Bogoliubov's next step rested on that the operators $a_0$ and
 $a^+_0$ of condensate atoms entered into the Hamiltonian in
 combination of $a_0/\sqrt{V}$ and $a^+_0/\sqrt{V}$ and within a
 large volume limit approximately commute with each other. At the
 same time, their matrix elements contain  $\sqrt{N_0}\,.$
 Therefore, the {\it operators $a_0$ and $a^+_0$ can be treated as
 numbers $\,=\sqrt{N_0/V}\,,$} and the operator ${\mathbf N}_0\,,$
 divided by $V\,,$ can be replaced by the finite density of Bose
 condensate $\rho_0=N_0/V\,.$  As a result, the Hamiltonian
 $H_{B_1}\,$  becomes a uniform bilinear form in operators with
 nonzero momentum
  \begin{equation}\label{BogHam2}
 H_{B_2}=\sum_{p\neq 0}\left\{[T(p)+\rho_0\,v(p)]\,b^+_p b_p\,+
 \tfrac{\rho_0\,v(p)}{2}\left[b^+_pb^+_{-p}+\,b_p b_{-p}\right]
 \right\}\,.\eeq
  It should be noted that the initial expression (\ref{BogHam0}),
 like (\ref{BogHam1}), is invariant with respect to phase
 transformation\footnote{By historical reasons transformation
 (\ref{phase}) is often called the gauge one, which might
 inevitably lead to association with \ ``electromagnetic gauge
 transformation''\ (as, e.g., in paper \cite{wein08}), i.e., with
 the law of electric charge conservation. This error was copied
 in the last Nobel press-release\cite{nobel-08}.} of the operators
 \begin{equation}\label{phase}
 a_k\to e^{i\phi}\,a_k\,,\quad a_k^+\to e^{-i\phi}\,a_k^+\,,\eeq
 which corresponds to conservation of particle number. Indeed,
 the Hamiltonian $H_{B_1}\,,$ like $H_{B_0}\,,$ commutes with
 the operator of total particle number
 ${\mathbf N}=\sum_k\,a^+_k a_k\,.$  However, this property is
 not inherent in approximation $H_{B_2}\,$ that does not
 contain condensate operators. Just this step, i.e., a
 transition to the bilinear (exactly solvable) approximate
 Hamiltonian (\ref{BogHam2}), leads to spontaneous symmetry
 breaking. \par

 The diagonalization of the bilinear Hamiltonian  $H_{B_2}\,$ is
 not a particular problem and can be accomplished by the famous
 Bogoliubov canonical $(u,v)$ transformation
  \begin{equation}\label{u-v}
  b_p \to \xi_p=u_p b_p+v_p b_{-p}^+\,;\quad b^+_p \to
  \xi^+_p=u_p b^+_p+v_p b_{-p}\,;\quad  u_p^2-v_p^2=1\eeq
 with real coefficients \ ``braiding'' \ creation and
 annihilation operators. Thus, the new operators
 $\xi_p\,$ and $\xi_p^+\,$ are a superposition of the old ones.
  A \ ``hyperbolic rotation'' \ of operators (\ref{u-v})
 corresponds to a unitary transformation\footnote{For technical
 details see, e.g., \S\ 12 and Appendix IV in text-book \cite{qf05}.}
 \begin{equation}\label{Ubog}
 b_p \to \xi_p=U^{-1}_{\alpha} b_p\,U_{\alpha}=
 u_p b_p+v_p b_{-p}^+\,;\quad U_{\alpha}= e^{\sum_p\alpha(p)
 \,[b_p^+b_{-p}^+ - b_p b_{-p}]}\,,\eeq
 where the coefficient $\alpha(p)$ depends on the parameters
 of the initial Hamiltonian. The transformed Hamiltonian is
 \begin{equation}\label{new_H}
 H=\sum_{p\neq 0}E(p)\,\xi_p^+\,\xi_p \eeq
 with the spectrum
 \begin{equation}\label{spectr}
 E(p)=\sqrt{T^2(p)+ T(p)\,v(p)\,\rho_0}\,. \eeq
 The new ground state
\begin{equation}\label{13}
\Psi_0(\alpha)=U_{\alpha}^{-1}\,\Phi_0=
 e^{-\sum_p\alpha(p)\,b_p^+b_{-p}^+}\,\Phi_0  \eeq
 includes superpositions of correlated pairs with the total zero
 momentum\footnote{It is interesting to note that a procedure
 similar to the Bogoliubov $(u,v)\,$ transformation is used
 (see, e.g. \cite{Yuen76}) in quantum optics in determining \
 ``squeezed'' \ states $\Psi_0(q)\sim \exp\{\sum_k a(k)\,[b_k^+
 \,b_{q-k}^+]\}\,\Phi_0\,,$ , where an important role is played
 by correlated pairs of photons with nonzero total momentum
 $q\,.$}. Transformation (\ref{u-v}), (\ref{Ubog}) leads to
 a spectrum of collective excitations  (\ref{spectr}). The
 dependence of energy on momentum has an initial linear part
 that is necessary for explanation of
 superfluidity and a nonlinear part with flexure that places
 Landau's rotons\footnote{The curve with flexure was published
 by Landau in article \cite{dau47} (also pp 32-34 in
 \cite{dau-2}) written soon after the
 discussion with Bogoliubov of his presentation of paper
 \cite{nn47} given on 21 October, 1946. In that article Landau
 used Bogoliubov's idea of a unique spectrum of collective
 excitations in quantum liquid. In a more detailed paper
 \cite{dau48} (see also pp 42-46 in \cite{dau-1} and
 \cite{dau49}) he emphasized Bogoliubov's priority: {\it
  ``It is worthwhile to point out that N.N. Bogoliubov has
 recently succeeded in determining in a general form an energy
 spectrum of Bose-Einstein gas with a weak interaction between
 particles with the help of ingenious application of the second
 quantization.''} \ Therefore, we think it appropriate to
 call the curve in Fig. 2[b] the Bogoliubov-Landau spectrum.}
 into a required position (see Fig. 2[b])
 \begin{figure}[ht!]
\begin{center} \label{fig2}
\includegraphics[scale=.80] {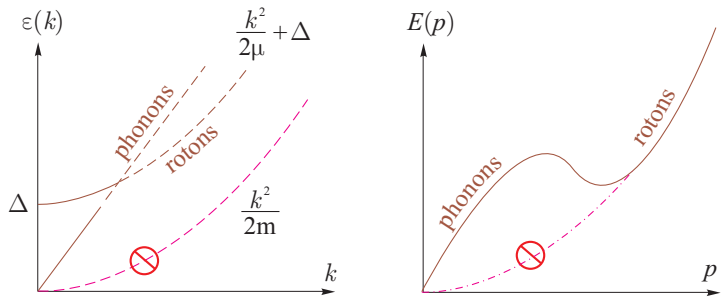}
 \caption{[a]\ Spectrum of phonons and rotons in the Landau
 phenomenological theory; \ [b]\ The Bogoliubov-Landau spectrum
 of collective excitations following from the expression
 (\ref{spectr}) of the Bogoliubov microscopic theory
 \cite{nn47,nn67}.}  \end{center}
 \end{figure}

 The absence of single-particle excitations, like in a
 phenomenological approach, underlies the formulation of the
 model, though an operator form of a canonical transfor\-mation
 gives information about the nature of collective excitations
 and the structure of the new ground state (\ref{13}).

 As mentioned above, the initial Hamiltonian of weakly non-ideal
 Bose gas (\ref{BogHam0}) is invariant with respect to gauge
 transformation  (\ref{phase}) providing conservation of the
 total particle number  $N\,.$ However, Bogoliubov's bilinear
 Hamiltonian (\ref{BogHam2}) has no this property, which
 corresponds to symmetry breaking. This Hamiltonian appeared
 as a result of the substitution of operator \ ``condensate'' \
 contributions (at  $k=0\,$)) by c-numbers. This substitution
 assumes nonzero values of vacuum averages $\langle a_0^+\rangle$
 and $\langle a_0\rangle\,,$ that are connected with a transition
 to the new vacuum by the unitary operator\footnote{
 See the footnote 8 above.}
 \begin{equation}\label{U-phase}
 U_{c}=e^{c\,(a_0^* - a_0)} \,,\quad a_q \to
 U_{c}^{-1}\,a_q\,U_{c} = b_q + c\,a_0\,.\eeq

  \subsection{Superconductivity}
  Another example of spontaneous symmetry breaking is the
 phenomenon of superconductivity where phase invariance
 violation occurs, as in the case of phase transition to a
 superfluid state. Though superconductivity was discovered
 in 1911, significantly earlier than $^4$He superfluidity, a
 theoretical insight into the phenomenon of superconductivity
 was gained much later than explanation of superfluidity. A
 breakthrough along this line was a phenomenological theory
 suggested by Ginzburg and Landau (G-L). In G-L theory
 \cite{G-L-50} a superconducting state was described by an
 effective ``wave function'' \ of superconducting electrons
 playing the role of a two-component order parameter
 \begin{equation}\label{gl-1}
 \Psi (\hbox{\bf r})=|\Psi ({\bf r})|\, {\rm exp}
 [i\Phi ({\bf r})] \, .\end{equation}

 The equilibrium properties of a superconductor are defined
 there by a free energy functional depending on
  $\Psi({\bf r})\,$  and external magnetic field
 ${\bf B}({\bf r})$:
 \begin{eqnarray} \label{gl-2}
 F(\Psi) &= & F_{n0}+\int d\,{\bf r}\left\{\frac{|{\bf B}|^2}
 {8\pi} + a|\Psi|^{2}+\frac{1}{2} b |\Psi|^4\right.\nonumber\\
 & + &\left.\sum_{\alpha}\frac{1}{2m^*}\left|\left(-i\hbar
 \nabla_{\alpha}-\frac{q}{c}\,A_{\alpha}\right)\Psi ({\bf r})
 \right|^2 \right\}\end{eqnarray}
 where $\,F_{n0}$ is free energy in a normal state,
 ${\bf B} = {\rm rot}\,{\bf A}\,,$  $\,q\,$ and $\, m^* $ are
 effective charge and mass of superconducting electrons. In the
 original paper, those were arbitrary parameters which on
 general physical grounds were put equal to electron charge
 and mass. Modulus of the order parameter (\ref{gl-1}) is
 proportional to the density of superconducting electrons
 $\,n_{s}\,,$ and its phase $\Phi({\bf r})$ defines a
 superconducting current
  \begin{equation} \label{S2a}
 j_{\alpha}=\frac{q\hbar}{m^*}\,|\Psi|^2\nabla_{\alpha}\,
 \Phi({\bf r})\,. \end{equation}

 The essential feature of the G-L theory is that at the
 temperature of a superconducting transition $T_{\rm c}$ the
 coefficient $\,a \sim (T - T_{\rm c})\,$ {\it changes the sign},
  while the positive coefficient $\,b\,,$ the effective mass
 $m^*\,$ and charge $q\,$ are independent of temperature. In
 such a case, the G--L functional (16) describes a transition
 from a normal state with $\Psi= 0$ to a superconducting one
 at $T = T_{\rm c}\,,$ at which a nonzero order parameter
 $\Psi\neq0\,$ arises. In the absence of a magnetic field,
 there occurs a second order phase transition with the
 mean-field critical indices. In the framework of the G-L
 theory, the behavior of a superconductor in an external
 magnetic field, including the Abrikosov vortex lattice in
 second type superconductors, was successfully described
 \cite{Abrikosov57}. At the same time, the nature of a
 superconducting transition remained unclear.

  We will comment on the structure of a \ ``potential''
  \ term in expression (16)
 \begin{equation}\label{V-gl}V(\varphi)= a\,
\varphi^2+\tfrac{b}{2}\,\varphi^4\,,\quad \varphi=|\Psi|\eeq

 \begin{figure}[ht!] \begin{center}  \label{fig3}
 \includegraphics[scale=.95]{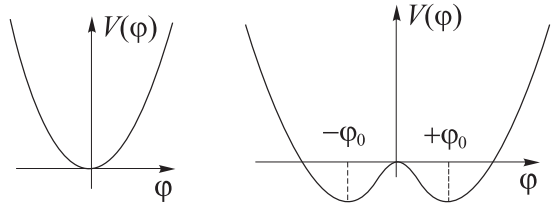}
 \caption{\fns [a]\ Potential energy of the free scalar field
 with mass $m^2>0\,;$ \ [b] Potential function of the scalar
 field with self-interaction and unstable symmetric state.}
 \end{center} \end{figure}
 \noindent in terms of a nonlinear (classical or quantum)
 oscillator. At $T > T_{\rm c}\,$ the coefficient $a\,$ is
 positive and can be expressed in terms of mass
 $a\to m^2/2\,.$ The first term dominates at small values of
 $\varphi\,$  and corresponds to an ordinary oscillator, like
 in Fig. 3[a]. Below the critical temperature this term is
 negative (see Fig. 3[b]) and the value $\varphi=0\,$
 becomes unstable, which results in spontaneous
 breaking of discrete symmetry related to reflection
 $\varphi\to- \varphi\,.$  Equation (\ref{V-gl}) and
 illustrations in Fig 3 correspond to a one-component order
 parameter. To the two-component case there correspond
 illustrations in Fig. 1 describing violation of continuous
 symmetry of rotation. \par
 The microscopic theory of superconductivity was developed
 only in 1957 by Bardeen, Cooper and Schiffer (BCS)
 \cite{BCS-57,BCS-57a} and Bogoliubov\cite{Bog58-1,Bog58nc}
 (see also \S 2 in \cite{BogBook58}, \cite{Bog58-2} and pp
 200-208 in \cite{nnVIII}). BCS considered a simplified model
 in which an interaction of electrons due to an exchange of
 phonons was substituted for an effective attraction of
 electrons near the Fermi surface
 \begin{eqnarray}
 H_{\rm BCS} =\sum_{{\bf k},\sigma}\,\varepsilon({\bf k})\,
 c^\dag_{{\bf k}\sigma} c_{{\bf k}\sigma} + \sum_{{\bf k},
 {\bf k}{'}} \,V_{{\bf k},{\bf k}{'}}\, c^\dag_{{\bf k}\uparrow}
  c^\dag_{-{\bf k} \downarrow} c_{-{\bf k}{'}\downarrow}
  c_{{\bf k}{'} \uparrow}\, , \label{bcs1} \\
 V_{{\bf k},{\bf k}^{'}}=\left\{\begin{array}{ll}-{V_{\rm BCS}},
 & |\varepsilon({\bf k}) - \varepsilon({\bf k}')|<\omega_{ph}\\
 0, & |\varepsilon({\bf k})-\varepsilon({\bf k}')|>\omega_{ph}\\
 \end{array} \right \}\nonumber\, ,
 \end{eqnarray}
 where $\, c^\dag_{{\bf k}\sigma} (c_{{\bf k}\sigma}) \, $
 are electron creation (annihilation) operators with momentum
 $\,{\bf k}\,$ and spin $\sigma =(\uparrow,\downarrow)=
 (+1/2, -1/2)\,,$ obeying the Fermi anticommutation relations:
 $$\,[c_{{\bf k}\sigma},c^{\dag}_{{\bf k'}\sigma'}]_+
 =\delta_{{\bf k, k'}}\delta_{\sigma,\sigma'}\,.$$
 The Bloch electron energy in the normal phase $\veps({\bf k})\,$
 is reckoned from the Fermi energy  $E_{\rm F}$, so that near the
 Fermi surface $\,\veps({\bf k})\approx{\bf v}_{\rm F}\cdot
 ({\bf k - k}_{\rm F}) \,$,  where $\,{\bf v}_{\rm F} =\partial
 \veps({\bf k})/\partial {\bf k} \,$ and $\,{\bf k}_{\rm F}\,$
 are the Fermi velocity and momentum,
 respectively. The interaction constant  $V_{\rm BCS}$
 defines attraction of electrons near the Fermi surface in a
 narrow energy layer $\,\pm\omega_{ph}\,,$ where
 $\,\omega_{ph}\,$ is a specific phonon energy. A variational
 wave function was used for calculation of the ground state
 energy and the spectrum of electron excitations.
  \begin{equation}\label{bsc3}
|\Psi{\rm BCS}\rangle=\prod_{{\bf k}}\left[\sqrt{1-h_{\bf k}}+
 \sqrt{h_{\bf k}}\,c^\dag_{{\bf k}\uparrow}\, c^\dag_{-{\bf k}
 \downarrow}\right]\,  |\Phi_0 \rangle  \, , \quad
 c_{{\bf k}\sigma}\, |\Phi_0 \rangle = 0\,,\end{equation}
 where the variational parameter $ h_{\bf k}],$ was determined
 from minimum of the ground state energy $\, W_0=\langle
 \Psi_{\rm\scs BCS}|H_{\rm BCS}|\Psi_{\rm BCS}\rangle\,.$
 It was established that an energy gap appears in the
 superconducting phase in the spectrum of one-electron
 excitations
  $$ \Delta \sim e^{-1/\lambda}\,;\quad
  E({\bf k})=\sqrt{\veps^2({\bf k})+|\Delta|^2}\, ,$$
 where the coupling constant $\lambda=V_{\rm BCS}\,N(0)\,$
 is determined by the effective interaction from Hamiltonian
 (\ref{bcs1}) and the density of electron states on the Fermi
 surface $N(0)\,.$ The thermodynamics and electrodynamics of
 a superconductor were considered, the temperature of a
 superconducting transition
 $\,T_c =1.14\,\omega_{ph}\,\exp\left(- 1/\lambda\right)$
 was calculated, and a universal relation between the gap in
 the spectrum at zero temperature and the temperature of a
 superconducting transition $\, 2\,\Delta_0 = 3.52\,T_c $ was
 obtained. The gap in the spectrum arises due to the formation
 of bound states of electron pairs with the opposite momenta
 and spins, \ ``Cooper pairs''. The corresponding vacuum
 expectation value
 \begin{equation} \label{S1}
 \langle c^{\dag}_{{\bf k}\uparrow}
 c^{\dag}_{{\bf -k}\downarrow}\rangle =\Psi({\bf k})=
 |\Psi( {\bf k})|\exp [i\Phi({\bf k})]\,\eeq
 represents an order parameter written in the form
 (\ref{gl-1}). This expression is explicitly related to
 the violation of phase (gauge) invariance
 \begin{equation}\label{S1a}
 c^{\dag}_{{\bf k}\uparrow }\rightarrow c^{\dag}_{
 {\bf k}\uparrow}\exp (i\varphi), \quad \langle c^{\dag}_{
 {\bf k}\uparrow}\,c^{\dag}_{{\bf -k} \downarrow} \rangle
 \rightarrow \langle c^{\dag}_{{\bf k}\uparrow}\,c^{\dag}_{
 {\bf -k}\downarrow}\rangle\,\exp(i 2\varphi)\end{equation}
 as in the theory of superfluidity \cite{nn47}. In this
 case, a long-range order in the superconducting phase is
 specified not only by the appearance of Cooper pairs,
 $\,|\langle c^{\dag}_{{\bf k}\uparrow}\,c^{\dag}_{{\bf
 -k}\downarrow}\rangle|\neq 0\,,$ but also by the fixation
 of the order parameter phase in the whole volume
 of a superconductor.\par
  Based on the BCS semi-phenomenological theory, Gor'kov
 \cite{Gorkov59} gave a consistent derivation of the G-L
 functional (\ref{gl-2}) and showed that an effective charge
 corresponds to a Cooper pair, i.e., $q= 2\,{\rm e}\,,$ and
 an effective mass should be taken equal to the mass of a
 Cooper pair $\;m^* =2\,m\,.$ In so doing, it is convenient
 to normalize the modulus of the order parameter to the
 density of superconducting electron pairs
 $\,|\Psi({\bf r})|^2 =n_{s}/2.$

 Before the appearance of a detailed BCS paper \cite{BCS-57a}
 Bogoliubov succeeded in construc\-ting a microscopic theory
 of superconductivity for the original Fr\"ohlich electron-phonon
 model
 \begin{equation}\label{S1b}
 H_{\rm Fr}= \sum_{{\bf k},\sigma}\,\varepsilon({\bf k})\,
 c^\dag_{{\bf k}\sigma} c_{{\bf k}\sigma} + \sum_{\bf q}
 \,\omega({\bf q})\, b^\dag_{{\bf q}} b_{{\bf q}}
 + g_{Fr}\sum_{{\bf k},{\bf q},\sigma}\,
 \sqrt{\tfrac{\omega({\bf q})}{2V}}\,
 c^\dag_{{\bf k}\sigma}c_{{\bf k +q}\sigma}( b^\dag_{{\bf q}}
 + b_{-{\bf q}})\,, \end{equation}
 where $\,\omega({\bf q})=s\,q\,,\,\,s\,$  is the velocity of
 sound, and the interaction of electrons with acoustic phonons
 is described by the  Fr\"ohlich coupling constant $\,g_{Fr}.$
 Generalizing the method of canonical $(u, v)$ transformation
 from the theory of superfluidity \cite{nn47,nn67} Bogoliubov
 introduced new Fermi amplitudes $\alpha_{{\bf k},\sigma}\,,$
 superpositions of electron creation and annihilation operators
 \cite{Bog58-1,Bog58nc}  (see also \S 2 in \cite{BogBook58}) :
\begin{equation} \label{uv-sc}
\alpha_{{\bf k}\uparrow}= u_{k}\,c_{{\bf k} \uparrow}-
 v_{k}\,c^{\dag}_{{-\bf k}\downarrow}\,\,, \quad
  \alpha_{{\bf k} \downarrow} = u_{k}\,c_{-{\bf k} \downarrow}
  +  v_{k}\,c^{\dag}_{{\bf k} \uparrow}\,\,;\quad
  u_{k}^2 + v_{k}^2 = 1\,,\end{equation}
 where $\,u_{k},\, v_{k}\,$  are real functions.

  The new Fermi amplitudes  $\,\alpha_{{\bf k} \sigma}\,$ and
 $\alpha^{\dag}_{{\bf k} \sigma}\,$  were used to carry out
 compensation of the so-called \ ``dangerous diagrams'' \
 responsible for the production of electron pairs with the
 opposite momenta and spins. In the Fermi amplitude
 representation (\ref{uv-sc}) the Hamiltonian of electrons
 in a superconducting state takes the form of Hamiltonian of
 the quasiparticle ideal gas
  \begin{equation} \label{B1}
 H_{\rm Fr} \to H_{\rm B} =\sum_{{\bf k},\sigma} E({\bf k})\,
 \alpha^{\dag}_{{\bf k}\sigma}\,\alpha_{{\bf k}\sigma}+ U_0,
 \quad E({\bf k}) =\sqrt{\varepsilon^2({\bf k})+
 |\Delta({\bf k})|^2},\eeq
 where the spectrum of excitations of quasiparticles
 $E({\bf k})$  is defined by the spectrum of electrons in the
 normal phase $\,\varepsilon({\bf k})\,$  and the gap in a
 superconducting state $\,\Delta({\bf k})\,$, depending on
 momentum k in the  general case. The equations derived by
 Bogoliubov for the gap and the superconducting temperature
 coincide with those in the BCS theory with the intensity
 directly determined by the Fr\"oehlich coupling constant
 in Hamiltonian (\ref{S1b}): $\,\lambda=g_{Fr}^2\,N(0)\,$.\par

 Bogoliubov's quasiparticles (\ref{uv-sc}) (sometimes called
 \ ``bogolons'') provide us with a clear physical picture of
 the spectrum of quasiparticle excitations as a superposition
 of a particle and a hole which have a gap in the spectrum on
 the Fermi surface. Let us give a spectral function of
 quasiparticle excitations in the superconducting phase
\begin{equation}\label{bog-spektr}                    
 A_{sc}({\bf k},\omega)={ u}_{\bf k}^2 \,\delta(\omega-
 E_{\bf k})+{v}_{\bf k}^2\;\delta(\omega+ E_{\bf k}), \eeq
 with due account for the expressions derived by Bogoliubov
 for the coefficients in transformation (\ref{uv-sc}):
 $$    u_{k}^2 =\frac{1}{2}\left[1 +\frac{\veps({\bf} k)}
 { E({\bf k})}\right],\quad v_{k}^2 = \frac{1}{2} \left[1 -
 \frac{\veps({\bf k})}{ E({\bf k})}\right ].$$
 Away from the Fermi surface, $|\veps({\bf k})|\gg|
 \Delta({\bf k})|,\;E({\bf k}) \approx |\veps({\bf k})|\,,$
 quasiparticle excitations are either electrons outside the
 Fermi sphere for $\,\veps({\bf} k)>0,\; u_{k}^2\approx
 1,\,v_{k}^2 \approx 0\, $, or holes inside the Fermi sphere
 for $\,\veps({\bf} k) < 0,\; u_{k}^2 \approx 0,\,v_{k}^2
 \approx 1\,$. In the vicinity of the Fermi surface,
 $|\veps({\bf k})|\ll|\Delta({\bf k})|,\;E({\bf k}) \approx
 |\Delta({\bf k})|$, excitations are a coherent superposition
 of an electron and a hole, so that the spectral function
 (\ref{bog-spektr}) has two peaks with equal weights:
  $\, u_{k}^2 \approx v_{k}^2 \approx 1/2\,$.  In this case,
 an energy gap for electron excitations equals
 $2 |\Delta({\bf k})|\,.$ When passing to the normal phase,
 $\,|\Delta({\bf k})| = 0\,,$ the spectral function takes a
 standard form $\,A_{n}({\bf k},\omega)= \delta(\omega -
 \veps({\bf} k))\,$  with a linear spectrum of excitations near
 the Fermi surface: $\,\veps({\bf k}) \approx {\bf v}_{\rm F}
 \cdot ({\bf k- k}_{\rm F}) \,$. Figure 4 shows the spectral
 phase of quasiparticle excitations (\ref{bog-spektr}) two
 branches of which correspond to the spectrum
 $\,\omega=\pm \,E({\bf k})$ with the gap $|\Delta({\bf k})|$.
\begin{figure}[th]\label{fig4}
\begin{center}
\includegraphics[scale=.23]{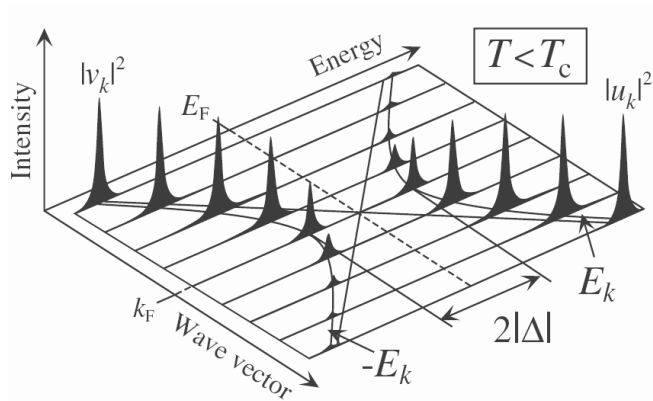}
\caption{Spectral function of one-electron quasiparticel
 exitations (\ref{bog-spektr}) of the Bogoliubov theory
 \cite{Bog58-1} in supercunducting phase (taken from
 paper~\cite{Matsui03}).}
\end{center}\end{figure} 

 A similar quasiparticle spectrum with two peaks was observed,
 for instance, in photoemission experiments in high-temperature
 superconductors \cite{Matsui03}), which proves the coherent
 nature of quasiparticles in the superconducting phase.

  Based on the Bogoliubov representation of quasiparticles it
 is easy to calculate thermodynamic and electrodynamic
 properties of a superconductor. The Bogoliubov canonical
 $(u, v)$ transformation (\ref{uv-sc}) is widely used in
 solving present-day problems in the theory of
 superconductivity.

 Noteworthy also is the generalization of the $(u, v)$
 transformation to the case of inhomogeneous systems, the
 Bogoliubov -- De Gennes transformation (see, e.g.,
 \cite{DeGennes66}) that can be written in terms of
 coordinate-dependent $(u({\bf r}), v({\bf r}))$ wave
 functions of electrons in the superconducting phase. Though
 the results obtained on the basis of the BCS simplified
 model Hamiltonian (\ref{bcs1}) were impressive, the problem
 of accuracy of the obtained solutions still remained
 unsolved. Omitting the details we should like to note
 that further analysis \cite{Bog-et-al57} (also pp 168-176
 in \cite{nnVIII}) showed that the superconducting phase
 represents a condensate of Cooper pairs (i.e., bosons)
 consisting of \ ``attracted'' \ electrons. The spectrum of
 excitations of a pair condensate satisfies the criterion
 of Landau's superfluidity. Thus, Bogoliubov came to the
 conclusion of the unity of these two phenomena: it is
 superfluidity of Cooper pairs that creates a superconducting
 current\footnote{Let us give a quotation from Bogoliubov's
 review paper \cite{Bog58c} (see also pp 289-296 in
 \cite{nnVIII}) of that time:\ {\it ``the property of
 superconductivity may be treated as a property of
 superfluidity of a system of electrons in metal''}.}. It
 should be pointed out that the identity of both these
 phenomena has recently been confirmed directly in
 experiments with ultracold fermion gases (see recent
 reviews \cite{bloch08,pitaev08}).

  Summing up the discussions of phase transitions in quantum
 statistics we should like to emphasize that in passing to the
 superfluid and superconducting state there occurs system's
 spontaneous symmetry breaking, namely, phase (otherwise gauge)
 invariance.

  \section{\sf SSB in quantum field theory}
 \subsection{\ns The 1960 events}
  First attempts to use the SSB mechanism in QFT arose in the year
 of 1960. At that time, this idea was as if in the air.
 Almost concurrently there appeared several investigations within
 two-dimensional (1+1) QFT models. The very first ones submitted
 for publication were papers by Vaks and Larkin \cite{v+l60}
 (also p 873 in  \cite{roch60}). More or less simultaneously,
 the first results by Tavkhelidze and Nambu \cite{nambu60}
 were obtained. That summer Bogoliubov and Nambu met in
 Utrecht\footnote{See the last sentence in paper
 \cite{utrecht60}.}, as well as three months later in
 Rochester at HEP Conference. There, Nambu delivered a draft
 of his first paper\footnote{Unfortunately, Nambu's
 publications contain only a slipshod reference to the
 preprint of the first Bogoliubov paper on superconductivity
 already published in JETP \cite{Bog58-1} two years before.
 Partially due to this, the appearance of the SSB phenomenon
 as early as 1946 in the Bogoluibov's theory of superfluidity
 \cite{nn47,nn67} (see pp 108-112 in \cite{nnVIII}) (as
 well as later on in the theory of superconductivity)
 relating to the (u,v) transformation and being physically
 responsible for the nonconservation of the number of
 noncondensed particles (of Cooper pairs) remained unnoticed
 by succeeding authors.} with Jona-Lasinio \cite{N-JL-60}.
 In the comment to the Nambu talk
 Bogoliubov said  (see page 865 in \cite{roch60})
\begin{quote}
  {\small\it ... one of my collaborators (Tavkelidze) has
 considered a Thirring--type one-dimensional model in which
 massless fermions interact with massive bosons. His calculations
 are not based on the self-consistent principle but on the
 ordinary Feynman diagram approach. The result is that there is
 a degeneracy in such a simple case.}
\end{quote}

  All the QFT models in the early papers \cite{v+l60} --
 \cite{N-JL-60} of 1960, including the most well-known second
 paper by Nambu and Jona-Lasinio \cite{N-JL-61}, were
 non-renormalizable, the results being dependent on cutoff.
 This drawback was avoided only by Arbuzov, Tavkhelidze and
 Faustov\cite{tav61} (see also pp 527-530 in \cite{nnIX}).

  The first successive use of the SSB took place several
 years later in the realistic model of electro-weak
 interaction by Glashow, Salam and Weinberg (GSW) where
 heavy gauge vector $W$ and $Z$ bosons acquire masses due
 to the Higgs mechanism.

 \subsection{\small The Higgs mechanism in Standard Model}
 Lagrangian of the complex (pseudo)scalar field with
 quartic self-interaction
 \[L(\varphi, g)= \frac{1}{2}\, \left(\partial_\mu\varphi
 \right)^2 -V(\varphi),\quad V(\varphi)=\frac{m^2}{2}\,
 \varphi^2+g\,\varphi^4;\quad g > 0 \]
 and stable lower state at $\varphi=0\,$ differs from
  Lagrangian of (two-component) Higgs field
 \begin{equation}\label{H-higgs}
  V_{\rm Higgs}(\Phi^2)=\lambda\left(\mathbf{\Phi}(x)^2-
 \Phi^2_0\right)^2  \,; \quad \mathbf{\Phi}^2= \Phi^2_1 +
 \Phi^2_2\,;\quad \Phi^2_0=\mbox{\small const.}\eeq
 by the sign of quadratic term\footnote{Cf. with Figs.
 3[a], 3[b] and with expression (\ref{V-gl}).}.

  This corresponds to pure imaginary initial mass
  $\mu^2_{\rm H}=-4\lambda\,\Phi^2_0\,.$ After the shift
 $\Phi_1(x)\to \varphi_1(x)=\Phi_1(x)-\Phi_0\,$ by a constant
 $\Phi_0\,$ there arises the physical mass of the Higgs field
 \begin{equation}\label{m-higgs}
 m_{\rm Higgs}=2\sqrt{2\lambda}\,\Phi_0\,,\end{equation}
 proportional to the vacuum expectation value $\Phi_0\,.$

  The main reason of this formal trick is the providing of
 nonzero masses to quanta of the above-mentioned gauge vector
 fields and to leptons and quarks. The first ones are expressed
 in terms of the coupling constants of the electro-weak interaction
 and $\Phi_0\,,$ like, e.g.,
  $M_Z=(e\,\sqrt{2}\,\Phi_0)/\sin 2\theta_W\,,$ while the last
 ones -- via  $\Phi_0\,$ and some Yukawa couplings. The Yukawa
 interactions involved are \ {\it especially added} \ to
 Lagrangian of the Standard Model for this and only this !
 purpose. These Yukawa interactions after the shift by
 $\Phi_0\,$
   $$ g_i \bar{\psi}_i\,\Phi(x)\,\psi_i\to  g_i \bar{\psi_i}\,
 \varphi(x)\,\psi_i+ m_i\,\bar{\psi_i}\,\psi_i;\,\quad m_i =
 g_i\,\Phi_0 $$
 provide masses for fermions.

  This recipe, devoid of elegance, gives masses to leptons and
 quarks at the barter rule  --
 {\sf one mass for one coupling constant}. As a result, of ca
 25 parameters (not counting neutrino masses) of the current
 Standard Model, just 12 are Yukawa constants, added \ ``by hand''.

  Nevertheless, in the gauge sector of the SM, the SSB phenomenon
 implemented in the form of the Higgs model, led, about 40 years
 ago, to one of the greatest triumphs of QFT -- prediction of
 the existence of neutral currents and numerical values of
 intermediate boson $W_{\pm}\,$ and $ Z_0\,$ masses.

 The 1979 Nobel Prize was awarded to theoreticians Glashow,
 Salam and Weinberg a few years before the experimental
 observation of $W_{\pm}\,$ and $Z_0\,$ particles, which, in
 its turn, was marked by another Nobel Prize in 1984.

  Along with quantum electrodynamics and quantum chromodynamics,
 the GSW theory of electro-weak interactions stands for a splendid
 achievement of the human intellect. Being based upon an elegant
 and powerful principle \ {\it ``dynamics from symmetry''}, it
 forms a foundation of the Standard Model.

  \subsection{\small Search of Higgs boson}
 Meanwhile, the v.a.v. $\Phi_0 \sim 250\,$ MeV, as defined from
 the electro-weak theory, is not sufficient for the estimation
 of the Higgs mass itself. Expression (\ref{m-higgs}) for the mass
 value contains also the self-interaction coupling $\lambda\,$
 that remains free. The current combination of theoretical and
 experimental restrictions results in a small window for
 possible mass value

 \centerline{114 \ GeV$\, < M_{\rm Higgs} < \ 154$ \ GeV,}
 \smallskip

 \noindent that, hopefully, quite soon has to be studied at the
 Large Hadron Collider.

 In the context of these \ ``great LHC expectations''\ it is worth
 reminding that a rather artificial Higgs construction (\ref{H-higgs})
 with its pure imaginary initial mass looks like a simple-minded
 relativistic replica of the Ginzburg--Landau classic functional
 (\ref{gl-2}), (\ref{V-gl}) with all its pragmatic advantages and
 physical shortcomings. The real underlying physical reason of SSB
 remains unknown, despite the electroweak theory success.

  In such a situation, any aspirations for direct experimental
 observation, in our opinion, look unjustifiably straightforward.

 \section{\sf Conclusion}
  \subsection{\small As regards practice of Nobel Committee
  on physics}

 Now a few words about the Nobel Prize awarding. The Committee
 on Physics is the Class for Physics of the Royal Swedish Academy
 consisting of six members. Just these Swedish Academicians
 take a decision along with the Alfred Nobel testament and
 taking into account opinions of the leading specialists, mainly
 the Nobel laureates community with its well-known specific
 features.

   Remind a few well known incidents. \par  \par

 Piotr Kapitsa discovered superfluidity in 1937. All his
 outstanding results in the low-temperature physics were obtained
 in the late 30s as well. He happened to be lucky enough to
 survive until his early nineties, when they bethought of him,
 more than 40 years later. We all remember in what a desperate
 state after an accident Landau got his Prize.\par

  Items of another kind. \par

  The 1999 Prize to t'Hooft and Veltman. The renormalization of
 the non-Abelian vector field, which acquired mass due to symmetry
 breaking, was a physically important and mathematically intricate
 problem. Its masterly solution by a rather complicated combination
 of formal tricks formed a base of electro-weak theory in late 60s
 and, subsequently, in Quantum Chromodynamics. However, the
 contribution of three Russian theorists to this solution is, at
 least, of no less importance than that of the laureates. Everyone
 calculates matrix elements (in electro-weak and QCD) by the
 Feynman rules formu\-lated by Faddeev and Popov and performs
 renormalization with due account of Slavnov's identities.

 Now the last case. It combines two important, but rather
 distinct from each other, elements of the Standard Model. Their
 junction seems rather deliberate. The first one, spontaneous
 symmetry breaking in the theory of quantum gauge
 fields, in the current context of the XXI century could be
 referred (a f t e r \ the Higgs particle discovery) to the
 names of Nambu, Goldstone and Higgs. The second one -- formal
 mixing of three lepton generations (via Cabbibo-Kobayashi-Maskawa
 matrix) in the current version of the Standard Model -- lies
 completely outside our scope.

   \subsection{\small Summary}

  Above, we attempted, in the fairy-tale form, to trace the
 development of a topical issue, spontaneous symmetry breaking,
 in the field of quantum physics during the XX century.   \par

  It is evident that the ``Nobel race'' is won by pragmatic
 theorists of the phenomenological kind, in terms of the
 introductory discussion. And this is natural, in a sense. Just
 in this mood [в этом ключе] the inventor of dynamite formulated
 the priority of benefit for people. A really clear-cut
 implementation of this spirit of the Nobel testament was the 2007
  Prize in physics.

  Meanwhile, the reductionists have no reasons to be dejected
 and envy. Their efforts' reward lies in other fields. Thanks
 to their achievements, a more complete picture of the physical
 universe appears; ties of affinity are established between
 unrelated, at first sight, phenomena such as between
 electromagnetic and nuclear forces and, quite hopefully, between
 dynamics of the Universe evolution and some hypothetic
 generalization of the Standard Model (with additional space
 dimensions).\medskip

  The author is indebted to Prof. Oleg Rudenko for the impetus of
 this talk and paper and continuous moral support. In the course
 of implementation of the initial plan, two mighty figures of
 Landau and Bogoliubov and complementary interference of their
 creative methods came to the fore. By a lucky chance, this
 paper is published just between their centennial jubilees. \medskip

 The role of Drs. N.M. Plakida and V.B. Priezzhev in composing
 Section 2 is indis\-pen\-sable. Practically, they are coauthors
 of it. Besides, they provided the author with a lot of subtle
 comments along the whole text. It was a pleasure to follow
 essential advices of Dr. V.A. Zagrebnov as well. This
 investigation was supported in part by the
 presidental grant Scientific School--1027.2008.2.
  
 \tableofcontents
 \end{document}